\documentclass[dvips]{acta}
\usepackage{supertabular,lscape,epsfig}
\usepackage{amssymb}
\usepackage{amsmath}
\SetPages{1}{10}
\SetVol{61}{2011}

\def \PASJ {{\it Publ.Astr.Soc.Japan}\/} 

\begin{document}

\begin{Titlepage}

\Title { Direct Evidence for Modulated Irradiation \\ 
         of Secondary Components in Dwarf Novae during Superoutbursts }

\Author {J.~~S m a k}
{N. Copernicus Astronomical Center, Polish Academy of Sciences,\\
Bartycka 18, 00-716 Warsaw, Poland\\
e-mail: jis@camk.edu.pl }

\Received{  }

\end{Titlepage}

\Abstract {  
Contribution from the irradiated secondary component is detected in  
the light curves of five dwarf novae observed during superoutbursts. 
Their superhump light curves show that irradiation is modulated with 
the superhump phase. 
This strengthens the new interpretation of 
superhumps (Smak 2009) as being due to the irradiation controlled mass transfer 
rate resulting in modulated dissipation of the kinetic energy of the stream. 
} 
{binaries: cataclysmic variables, -- Stars: dwarf novae }

\section { Introduction } 

From the analysis of the superoutburst light curves of four deeply eclipsing dwarf 
novae it was found (Smak 2009,2010) that the observed brightness of the disk 
is modulated with phase of the beat period. The amplitudes of this modulation 
and the phases of maximum obtained for the four systems were practically identical, 
their average values being: $<A>=0.18\pm 0.01$ and $<\phi_b^{max}>=0.65\pm 0.02$. 
Similar modulation was also detected in superhump amplitudes of six dwarf 
novae (Smak 2010), with values of $\phi_b^{max}$ being less accurate 
($0.62\pm 0.06$ or $0.68\pm 0.08$) but consistent with the original value. 
This was interpreted as being due to a non-axisymmetric structure of the disk, 
involving the azimuthal dependence of its geometrical thickness, rotating 
with the beat period. 

Regardless of this interpretation it is obvious that modulation with the beat phase  
seen by the observer translates into modulated irradiation of the secondary 
component with the superhump phase. In particular, that the maximum irradiation 
of the secondary occurs at $\phi_{sh}^{max}=-\phi_b^{max}=0.35\pm 0.02$. 

This became one of the crucial ingredients of the new interpretation of 
superhumps (Smak 2009) in terms of the irradiation controlled mass transfer 
rate resulting in modulated dissipation of the kinetic energy of the stream. 

The purpose of this paper is to present results of an analysis of superoutburst 
light curves of five dwarf novae which provide {\it direct} evidence for the modulated 
irradiation of the secondary, thereby strengthening this interpretation. 
In Section 2 estimates are presented showing that 
the expected contribution from the irradiated secondary to the total brightness 
of the system should be detectable. 
The data and the methods of their analysis are described in Sections 3 and 4. 
In Section 5 the composite orbital light curves at superhump phases around 
$\phi_{sh}^{max}$ are presented showing the expected dependence on $\phi_{orb}$. 
In Section 6 the composite superhump light curves near $\phi_{orb}=0.5$ are compared 
to those near $\phi_{orb}=0.0$, showing the expected modulation with $\phi_{sh}$.

\section { The Irradiated Secondary Component }

We begin by estimating the expected contribution from the irradiated secondary 
to the total brightness of the system. To calculate its absolute visual magnitude 
we proceed as follows. The distribution of temperature over the irradiated 
hemisphere of the secondary is described by      

\beq
\sigma T^4~=~{{L_{BL}}\over {4~\pi~D^{~2}}}~\cos \theta~,
\eeq

\noindent
where $D$ is the distance from the primary to the point considered, 
$\theta$ is the incidence angle and  

\beq
L_{BL}~=~{{1}\over {2}}~{{G~M_1}\over {R_1}}~\dot M~
\eeq

\noindent
is the luminosity of the boundary layer. 

The absolute visual magnitude of the irradiated secondary is 
then caclulated by integrating the emerging flux over its irradiated portion,  
i.e. excluding the equatorial parts which are in the shadow cast by the disk. 
The resulting magnitudes $M_{V,2}(\dot M,z/r)$ are functions 
of the accretion rate and of the disk thickness parameter $z/r$. 
Then the magnitudes of the system with and without irradiation are calculated, 
their difference $\Delta M_V(\dot M,z/r)$ giving the contribution from the irradiated 
secondary to the total brightness of the system. 
We adopt system parameters of Z Cha (Smak 2007), accretion rates between 
$\dot M=1\times 10^{17}$ and $3\times 10^{17}$ (as observed in Z Cha during its 
superoutbursts; Smak 2008a), and $z/r=0.05-0.20$.  
The resulting values of $\Delta M_V$ depend primarily on $z/r$, ranging from 
$\Delta M_V\sim 0.45$ at $z/r=0.05$ to $\Delta M_V\sim 0.25$ at $z/r=0.20$.

\section { The Data }

The data to be used in our analysis, consisting of light curves of dwarf 
novae observed during their superoutbursts, were selected from the literature 
using two requirements: (1) a good coverage in orbital and superhump phases,  
and (2) the superhump amplitudes larger than $\sim 0.10$ mag. 
Sufficient data were found for the following five stars: 

{\it OY Car}. Light curves observed by Krzemi{\'n}ski and Vogt (1985; Fig.2b) 
on January 4-9, 1980, covering 17 superhumps.  

{\it Z Cha}. Light curves observed during several superoubursts by 
Warner and O'Donoghue (1988; Fig.1), with orbital phases of superhump maxima 
listed in their Table 3; regretfully, some of those phases are rather uncertain. 

{\it XZ Eri}. Light curves observed by Uemura et al. (2004; Fig.2) on JD 2452668, 
669, and 670, covering 8 superhumps. 

{\it VW Hyi}. Three sets of data consist of light curves observed by 
Vogt (1974; Fig.5) on December 16-21, 1972; by Verbunt et al. (1987; Fig. 1); 
and by Schoembs and Vogt (1980; Fig.2a, Runs 1 and 2). 
VW Hyi is a non-eclipsing system, but its orbital period is well determined 
from "hot spot humps" observed at quiescence. We use the elements from 
van Amerongen et al. (1987) corrected by $\Delta \phi_{orb}=0.15$ (Smith et al. 2006) 
accounting for the delay between the spectroscopic conjunction and the hump maximum.   

{\it DV UMa}. Light curves observed by Patterson et al. (2000; Fig.4) on 
JD 2450553 and 554, covering 12 superhumps. 

The data to be analysed in further sections consist of magnitudes read 
from the published light curves at specific superhump phases $\phi_{sh}$ (see Section 5) 
or at specific orbital phases $\phi_{orb}$ (see Section 6).  
In each case the corresponding orbital phase or superhump phase is also determined. 
In effect we have $m(\phi_{orb},\phi_{sh})$. 
In addition, the magnitudes at maximum ($m_{max}$) and at minimum ($m_{min}$) of 
a given superhump cycle are obtained and used to determine the local values of the 
superhump amplitude: $A_{sh}=m_{min}-m_{max}$ and of the mean magnitude: 
$<m>=(m_{min}+m_{max})/2$. 

Prior to further analysis the observed light curces must be corrected for 
the presence of the "hot spot hump". In the case of Z Cha it was found (Smak 2007) 
that such a hump, with amplitude $A_{spot}\approx 0.15$ and maximum at 
$\phi_{orb}^{max}\approx 0.95$, is present when the corresponding beat phase 
$\phi_b=\phi_{orb}-\phi_{sh}$ is between $\sim 0.3$ and $\sim 0.7$ and absent 
at other beat phases; identical values of $A_{spot}$ and $\phi_{orb}^{max}$ 
were obtained also for OY Car (Smak 2008b). 
Using this evidence we correct the observed magnitudes corresponding to 
$\phi_b=0.3-0.7$ and -- simultaneously -- to $\phi_{orb}=0.70-0.20$ by 

\beq
\delta m_{spot}~=~0.15~\cos (\phi_{orb}-0.95)~. 
\eeq

It is known that the brightness of a dwarf nova and the amplitude of its superhumps 
decrease during superoutburst, showing also some additional variations. 
To compensate for those effects we define the reduced magnitudes as 

\beq
\Delta m(\phi_{orb},\phi_{sh})~=~\left [~m(\phi_{orb},\phi_{sh})~-~<m>\right ]
       {{<A_{sh}>}\over {A_{sh}}}~, 
\eeq 

\noindent
where $<A_{sh}>$ is the average superhump amplitude obtained from all data for 
a given star. 

From considerations in the next Section it is clear that the effects of modulated 
irradiaton of the secondary modify the shape of the entire superhump light curve 
including $m_{min}$ and $m_{max}$. Consequently the values of 
$\Delta m(\phi_{orb},\phi_{sh})$ are subject to uncertainties introduced 
by the use of the specific "prescription" defined by Eq.(4). 
We shall return to this problem in Section 6.

\section { Qualitative Condiderations }

The observed brightness of an irradiated secondary depends on the orbital 
phase, with maximum at $\phi_{orb}^{max}=0.5$, i.e. when its irradiated hemisphere 
is facing the observer. It also depends on the orbital inclination, with maximum 
effect to be seen at $i=90^\circ$. 
On the other hand, according to the evidence already available (see Introduction), 
the irradiation is modulated with the superhump phase, with maximum at 
$\phi_{sh}^{max}\approx 0.35$. 

If so, the effects of modulated irradiation should be visible 
in the observed light curves in the form of excess brigtness near $\phi_{sh}^{max}$ 
when it coincides with $\phi_{orb}=0.5$. 
Such an excess can actually be seen in some of the light curves. For example:  
in OY Car on January 7 and 8, 1980 (Krzemi{\'n}ski and Vogt 1985; Fig.2b), 
in Z Cha during run S2740 (Warner and O'Donoghue 1988; Fig.1), 
in XZ Eri on JD 2452668 and 669 (Uemura et al. 2004; Fig.2), 
and in DV UMa on JD 2450554 (Patterson et al. 2000; Fig.4). 

The problem is, however, that irradiation effects can be seen in the {\it observed} 
light curves only selectively -- in narrow intervals of phases. 
An obvious alternative is to construct {\it composite} light curves consisting 
of points corresponding to the specific orbital or superhump phases. 
Accordingly, we adopt the following two-step strategy for our analysis. 
First, in order to check whether the irradiation effects are really 
present and detectable, we will construct the {\it composite orbital} light curves 
using points with superhump phases near $\phi_{sh}^{max}$. 
Secondly, in order to detect the effects of variable irradiation, we will construct 
the {\it composite superhump} light curves using points with orbital phases near 
$\phi_{orb}\sim 0.5$, i.e. close to the maximum irradiation, and compare them 
to the superhump light curves corresponding to $\phi_{orb}\sim 0.0$, 
i.e. to the phase where irradiation effects are absent.

\section { The Orbital Light Curves -- Detection of Irradiated Secondary } 

As discussed above the expected contribution from the irradiated secondary 
should be best seen in the composite orbital light curves corresponding 
to $\phi_{sh}^{max}\sim 0.35$. 
To account for the uncertainty of this value and to increase the number of points 
we construct the orbital light curves using $\Delta m(\phi_{orb},0.3)$ 
and $\Delta m(\phi_{orb},0.4)$. 

The resulting {\it composite orbital} light curves for four stars are shown in Fig.1 
(XZ Eri could not be included because of insufficient coverage in $\phi_{orb}$).  
As can be seen, those light curves show the expected dependence on $\phi_{orb}$ 
quite clearly. The best fit cosine curves are shown in those figures mainly 
for illustrative purposes; in view of the large scatter and for reasons discussed 
in the next Section not much weight can be given to their parameters and for 
that reason they are not listed here. 

\begin{figure}[htb]
\epsfysize=8.0cm 
\hspace{1.5cm}
\epsfbox{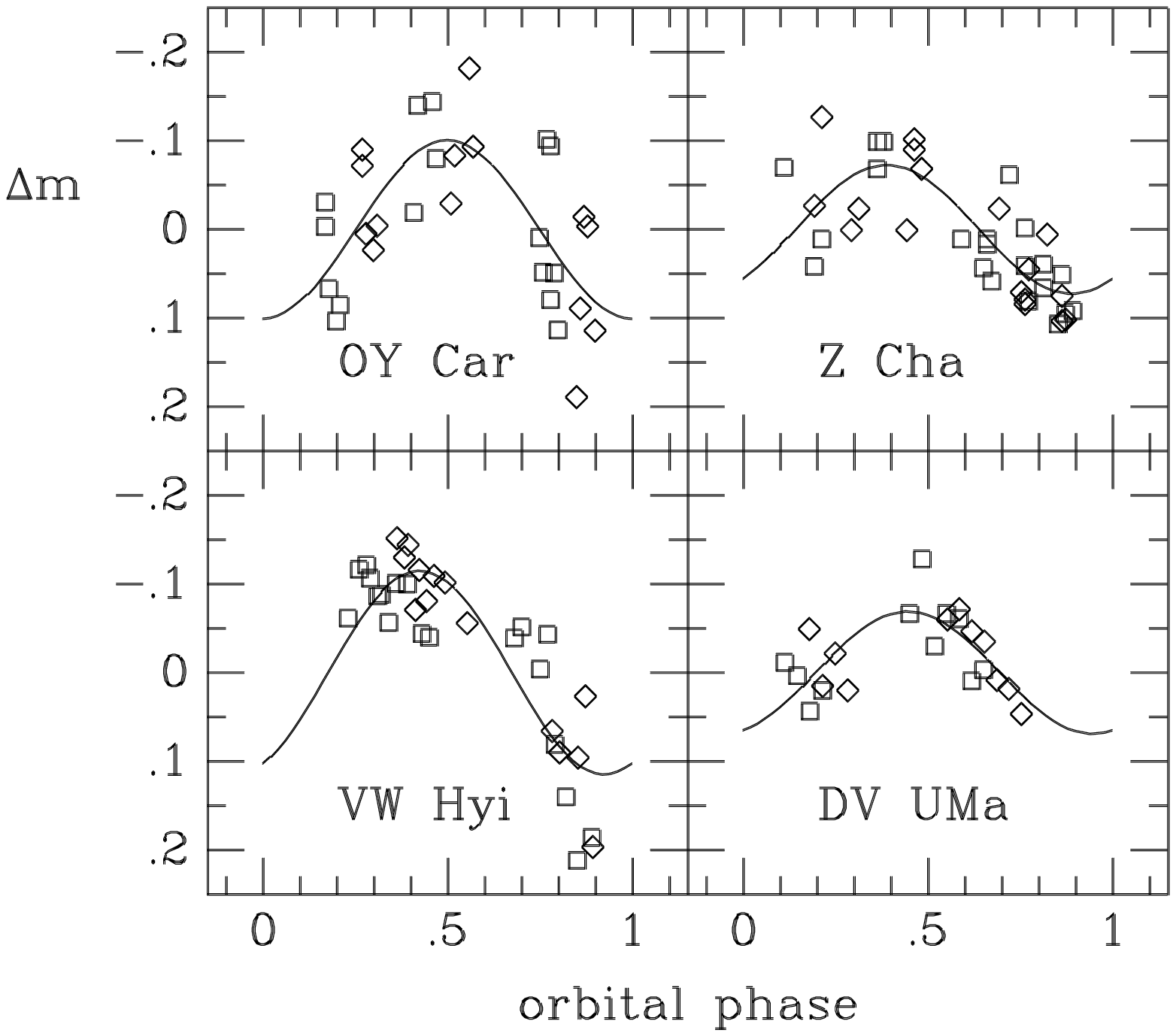} 
\vskip 5truemm
\FigCap { The composite orbital light curves of four dwarf novae at $\phi_{sh}=0.3$ 
({\it squares}) and $\phi_{sh}=0.4$ ({\it diamonds}). Best fit cosine curves are 
also shown.  
} 
\end{figure}

There is one problem which requires comments. The light curves of high inclination 
systems should show a secondary eclipse near $\phi_{orb}=0.5$ due to partial 
occultation of the secondary by the disk. It can be estimated, however, that 
the central depth of this eclipse is very low -- comparable to the scatter of points 
in the light curves in Fig.1. To confirm those estimates we review the original 
light curves and find that such a very shallow eclipse is indeed present 
in only some of them: in the January 6, 1980, light curve of OY Car 
(Krzemi{\'n}ski and Vogt 1985; Fig.2b) and in the S0130 and S2740 light curves of 
Z Cha (Warner and O'Donoghue 1988; Fig.1).

\section { The Superhump Light Curves -- Detection of Modulated Irradiation }

As suggested in Section 4 the expected contribution from the irradiated 
secondary should be best seen in the superhump light curve around $\phi_{orb}=0.5$ 
and absent around $\phi_{orb}=0.0$. 
Accordingly we determine the {\it composite superhump} light curves at $\phi_{orb}=0.4$, 
0.5 and 0.6 and at $\phi_{orb}=0.9$ and 0.1 (for VW Hyi we also use $\phi_{orb}=0.0$). 

Starting with superhump light curves near $\phi_{orb}=0.0$, shown in Figs.2a-6a, 
we note that they are fairly smooth and can be fitted with a simple cosine shape: 

\beq
\Delta m_{sh}^\circ (\phi_{sh})~=~\Delta m_\circ~+~A_{sh}~\cos \phi_{sh}~.  
\eeq

Turning to superhump light curves near $\phi_{orb}=0.5$, shown in Figs.2b-6b, 
we note that they differ systematically from those near $\phi_{orb}=0.0$, 
showing the expected excess brightness around $\phi_{sh}\sim 0.5$. 
To show it more clearly we calculate the residuals  

\beq
\delta m(0.4-0.6,\phi_{sh})~=~
\Delta m(0.4-0.6,\phi_{sh})~-~\Delta m_{sh}^\circ (\phi_{sh})~, 
\eeq

\noindent
where $\Delta m_{sh}^\circ (\phi_{sh})$ is given by Eq.5.

\begin{figure}[htb]
\epsfysize=8.5cm 
\hspace{2.5cm}
\epsfbox{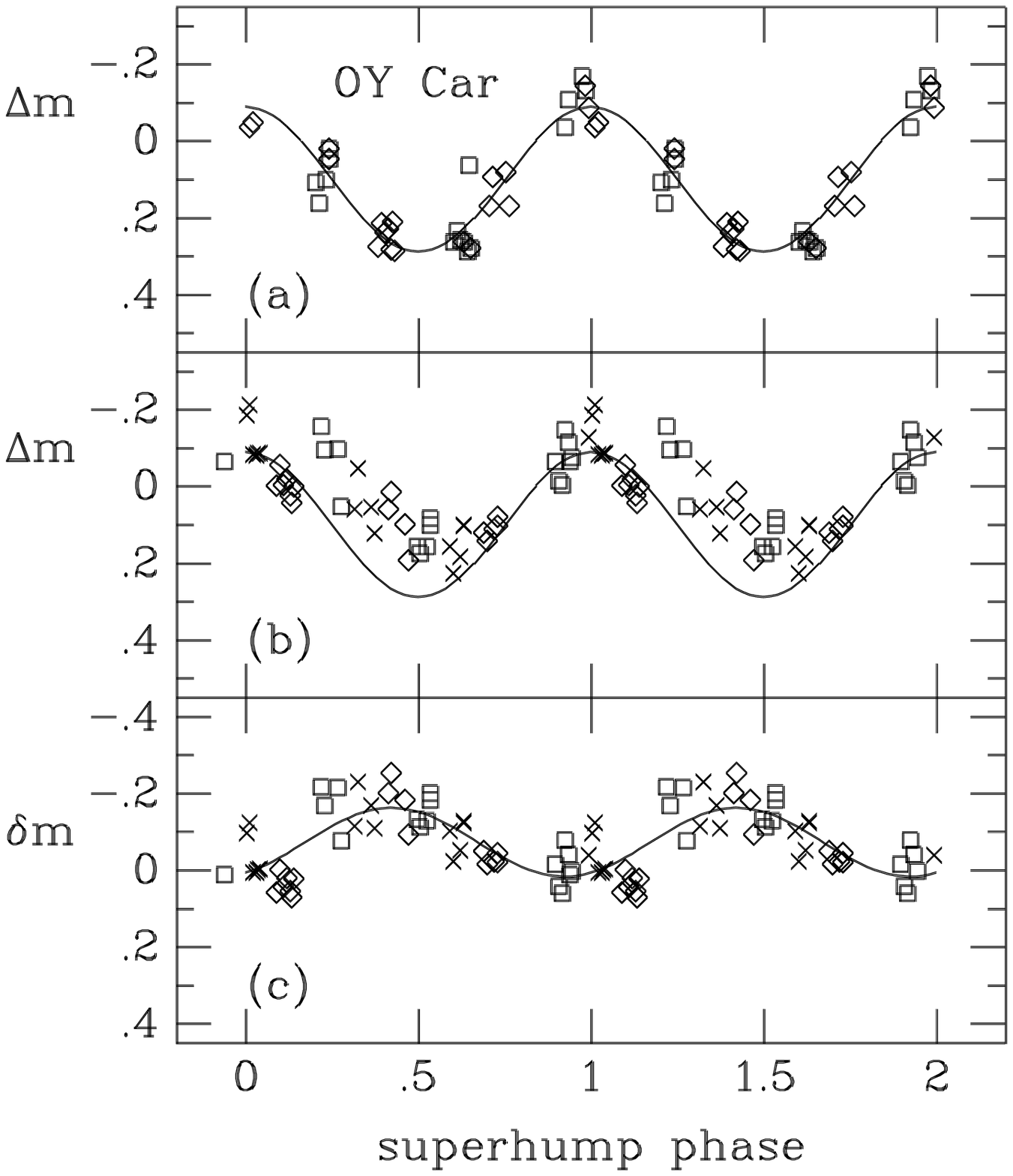} 
\vskip 5truemm
\FigCap { (a) The composite superhump light curve of OY Car at $\phi_{orb}=0.1$ 
({\it squares}) and $\phi_{orb}=0.9$ ({\it diamonds}). Solid line is the best fit 
cosine curve. 
(b) The composite superhump light curve of OY Car at $\phi_{orb}=0.4$ ({\it squares}), 
$\phi_{orb}=0.5$ ({\it crosses}) and $\phi_{orb}=0.6$ ({\it diamonds}). 
Solid line is the same as in (a). 
(c) The residuals (see Eq.6) between points in (b) and the best fit cosine curve. 
Symbols are the same as in (b). Solid line is the best fit cosine curve representing 
the residuals. 
} 
\end{figure}

\begin{figure}[htb]
\epsfysize=8.5cm 
\hspace{2.5cm}
\epsfbox{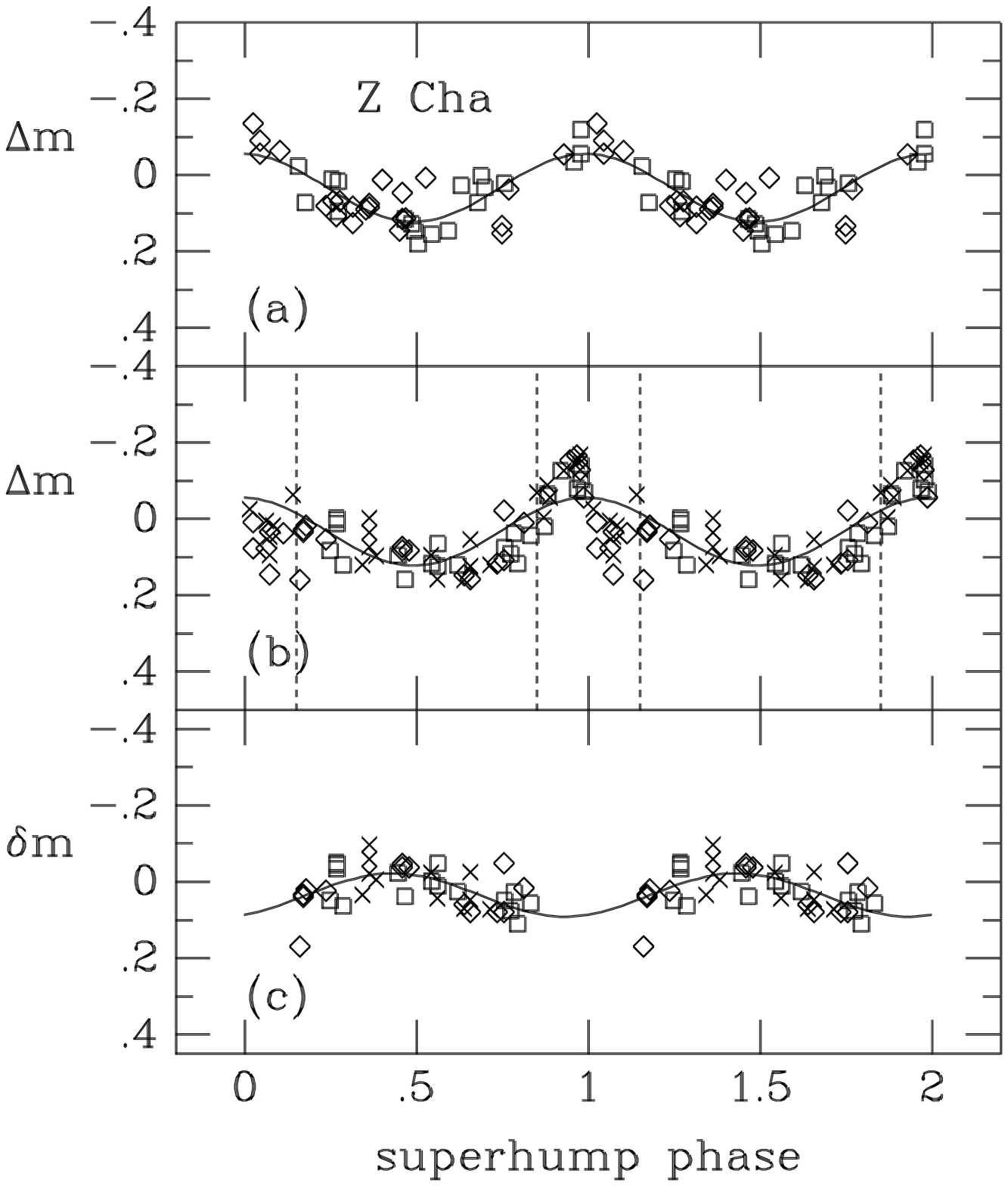} 
\vskip 5truemm
\FigCap { The composite superhump light curves of Z Cha. For explanations -- 
see caption to Fig.2. Dotted lines in (b) show the interval of phases not used 
in the analysis of residuals shown in (c). See text for details. 
} 
\end{figure}

\begin{figure}[htb]
\epsfysize=8.5cm 
\hspace{2.5cm}
\epsfbox{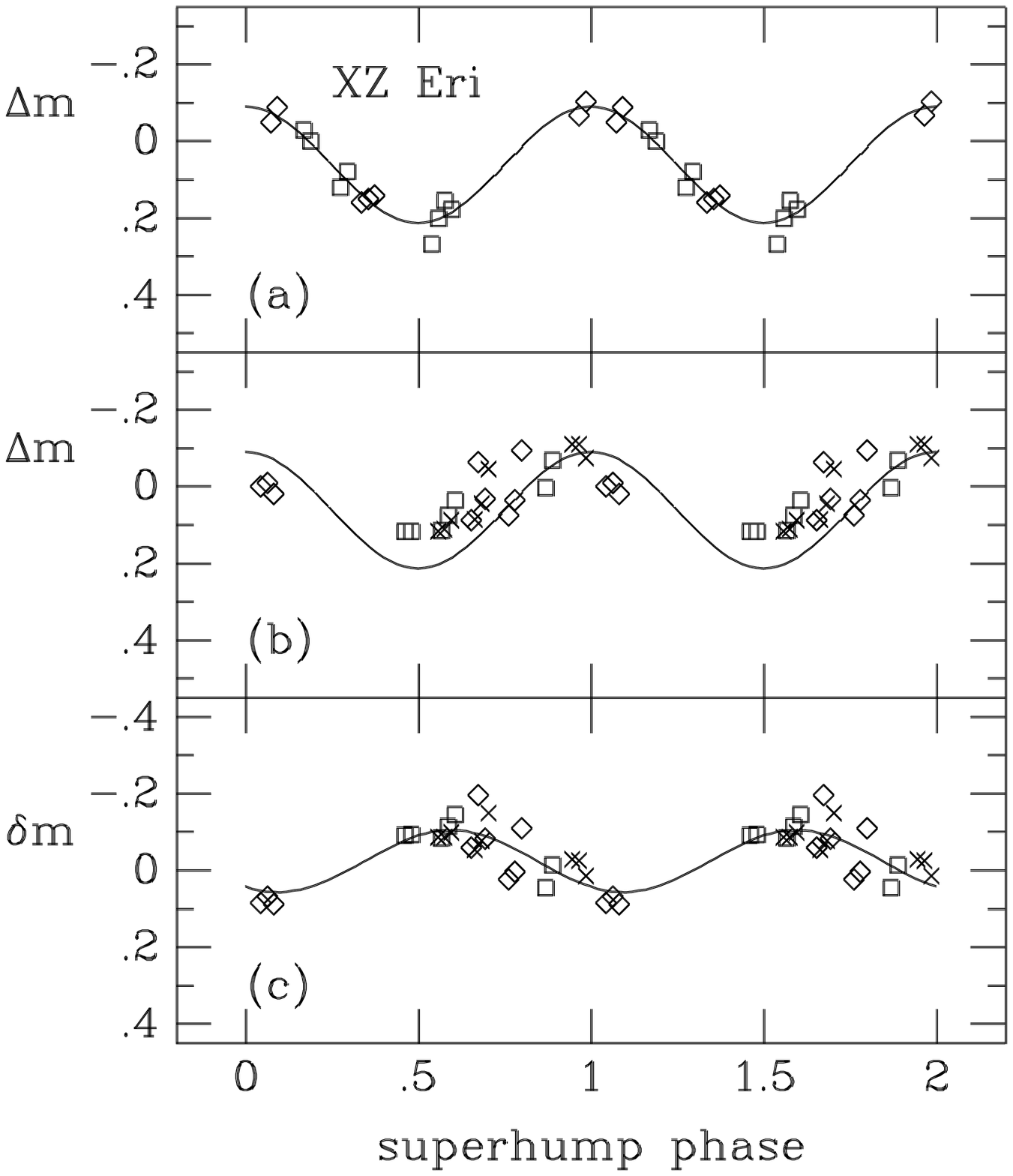} 
\vskip 5truemm
\FigCap { The composite superhump light curves of XZ Eri. 
For explanations -- see caption to Fig.2. 
} 
\end{figure}

\begin{figure}[htb]
\epsfysize=8.5cm 
\hspace{2.5cm}
\epsfbox{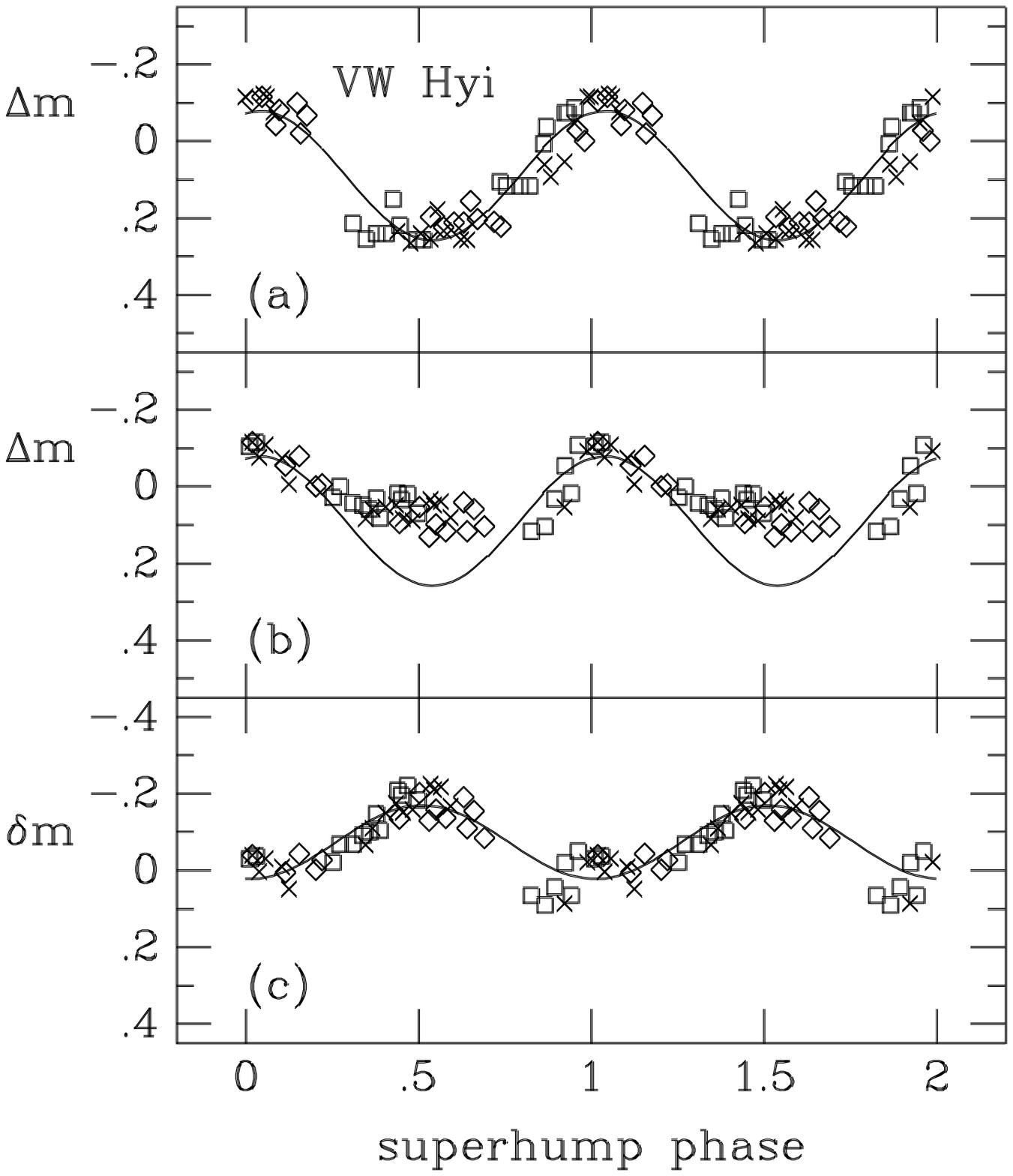} 
\vskip 5truemm
\FigCap { The composite superhump light curves of VW Hyi. Crosses in (a) correspond 
to $\phi_{orb}=0.0$. For other explanations -- see caption to Fig.2. 
} 
\end{figure}

\begin{figure}[htb]
\epsfysize=8.5cm 
\hspace{2.5cm}
\epsfbox{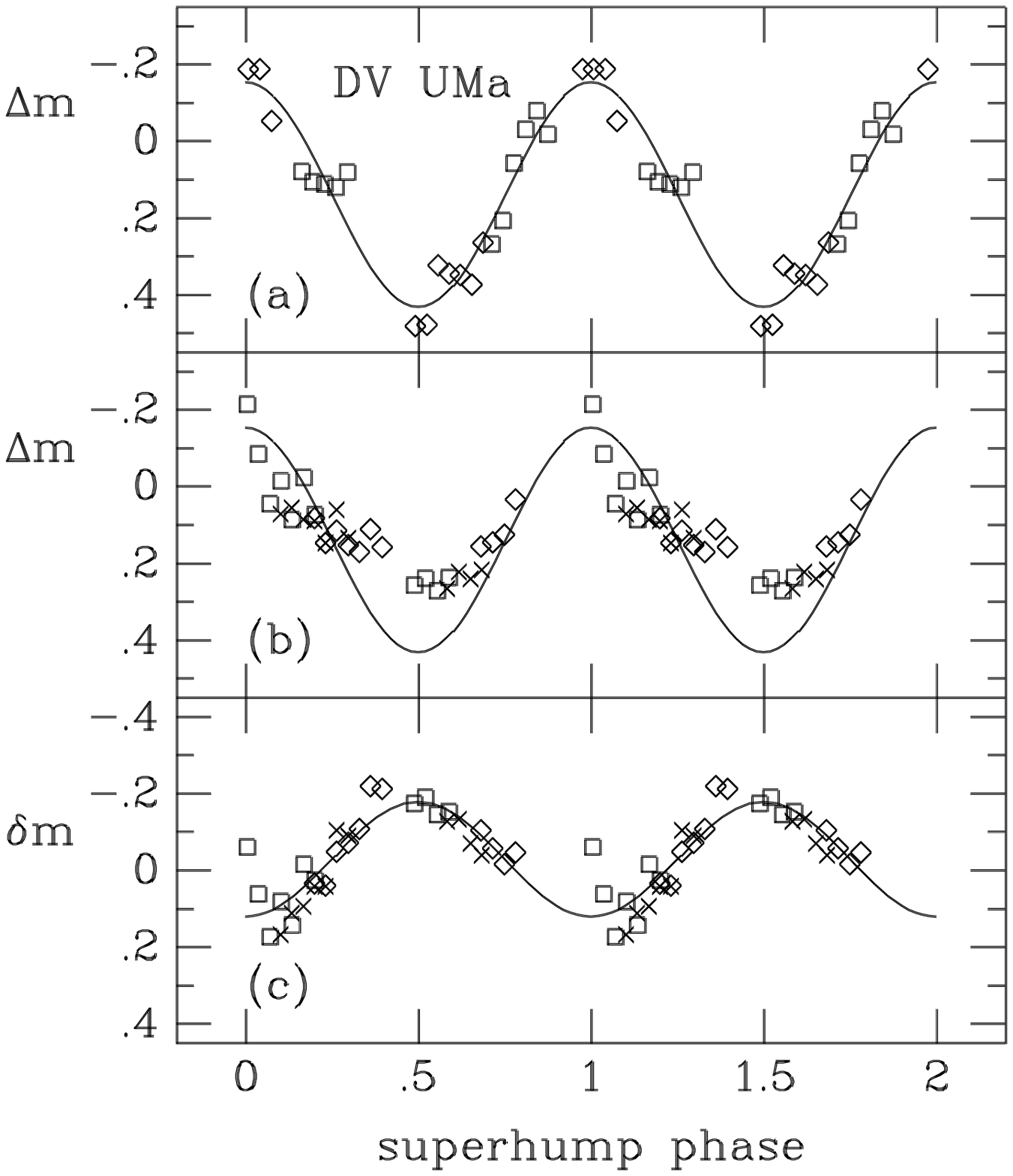} 
\vskip 5truemm
\FigCap { The composite superhump light curves of DV UMa. 
For explanations -- see caption to Fig.2. 
} 
\end{figure}

Those residuals, plotted in Figs.2c-6c, clearly show the expected modulation 
with $\phi_{sh}$. Shown also are the best fit cosine curves, their parameters 
$A_{irr}$ and $\phi_{sh}^{max}$ being listed in Table 1. 

\begin{table}[h!]
{\parskip=0truept
\baselineskip=0pt {
\medskip
\centerline{Table 1}
\medskip
\centerline{ Parameters of Modulated Irradiation }
\medskip
$$\offinterlineskip \tabskip=0pt
\vbox {\halign {\strut
\vrule width 0.5truemm #&	
\quad\hfil#\hfil\quad&	        
\vrule#&			
\quad\hfil#\hfil\quad&          
\vrule#&			
\quad\hfil#\hfil\quad&	        
\vrule width 0.5 truemm # \cr	
\noalign {\hrule height 0.5truemm}
&&&&&&\cr
&  Star  &&$A_{irr}$	  &&$\phi_{sh}^{max}$ &\cr
&&&&&&\cr
\noalign {\hrule height 0.5truemm}
&&&&&&\cr
& OY Car &&$ 0.09\pm 0.02$&&$0.42\pm 0.03$&\cr
&  Z Cha &&$ 0.06\pm 0.02$&&$0.43\pm 0.04$&\cr
& XZ Eri &&$ 0.08\pm 0.02$&&$0.60\pm 0.04$&\cr
& VW Hyi &&$ 0.10\pm 0.01$&&$0.51\pm 0.02$&\cr
& DV UMa &&$ 0.15\pm 0.02$&&$0.50\pm 0.02$&\cr
&&&&&&\cr
\noalign {\hrule height 0.5truemm}
}}$$
}}
\end{table}

The case of Z Cha requires special comment. Fig.3b shows  
peculiar distribution of points near $\phi_{sh}=0.0$. This is most likely 
due to very narrow shapes of some of the superhumps and uncertain phases of their 
maxima. Rather than arbitrarily modifying those points we simply exclude them 
from the analysis of residuals shown in Fig.3c. 

We now return to the problem of uncertainties related to the use of the specific 
form of Eq.(4) adopted in Section 3. 
To clarify this point we repeated our analysis using two alternative versions of  
this equation, with $<m>$ being replaced by $m_{min}$ or $m_{max}$, and its 
three other versions without factor $<A_{sh}>/A_{sh}$. 
Results were {\it qualitatively} the same, showing clearly the effects of modulated 
irradiation. As could be expected, however, the amplitudes $A_{irr}$ 
turned out to be sensitive to the particular form of Eq.(4); therefore 
their values listed in Table 1 should not be given much weight. On the other hand, 
the phases of maximum $\phi_{sh}^{max}$ depend only slightly on the particular form 
of Eq.(4).

\section { Conclusions } 

The composite superhump light curves presented in Section 6 clearly show 
that the irradiation of the secondary component is modulated with the superhump 
phase. Subject to uncertainties discussed in that Section, the full amplitude of 
those variations $<2A_{irr}>\approx 0.2$ mag (which is actually the 
difference between maximum and minimum irradiation) is consistent 
with estimates presented in Section 2. 

The mean phase of maximum irradiation $<\phi_{sh}^{max}>=0.49\pm 0.03$ for the 
five stars listed in Table 1 differs from $<\phi_{sh}^{max}>=0.35\pm 0.02$ 
obtained earlier from indirect evidence (see Introduction). 
The origin of this difference is quite obvious: 
the value $<\phi_{sh}^{max}>=0.35$ refers to the moment of maximum irradiation, 
while $<\phi_{sh}^{max}>=0.49$ -- to the moment when its consequences show up 
in the light curve. In other words -- the difference $\Delta \phi_{sh}\approx 0.14$ 
represents the response time of the atmospheric layers to variable irradiation. 
Model calculations by Hameury et al. (1988, Fig.1) show that the transition 
from the initial state without irradiation to the state when a hot isothermal 
layer extends to sub-photospheric layers occurs on a time scale longer 
than $\Delta t\sim 10^3$s. 
Once such a layer is formed, however, its response time to variable irradiation 
becomes shorter. In our case we have: $\Delta t=\Delta \phi_{sh} P_{sh}\approx 900$s.


\begin {references} 


\refitem {Hameury, J.M., King, A.R., Lasota, J.P.} {1988} {\AA} {192} {187}

\refitem {Krzemi{\'n}ski, W. and Vogt, N.} {1985} {\AA} {144} {124}

\refitem {Patterson, J., Vanmunster, T., Skillman, D.R., Jensen, L., 
          Stull, J., Martin, B., Cook, L.M., Kemp, J., Knigge, C.} 
          {2000} {\PASP} {112} {1584} 

\refitem {Schoembs, R. and Vogt, N.} {1980} {\AA} {91} {25}

\refitem {Smak,J.} {2007} {\Acta} {57} {87}    

\refitem {Smak,J.} {2008a} {\Acta} {58} {55} 

\refitem {Smak,J.} {2008b} {\Acta} {58} {65} 

\refitem {Smak,J.} {2009} {\Acta} {59} {121} 

\refitem {Smak,J.} {2010} {\Acta} {60} {357}  

\refitem {Smith, A.J., Haswell, C.A., Hynes, R.I.} {2006} {\MNRAS} {369} {1537} 

\refitem {Uemura, M. et al.} {2004} {\PASJ} {56} {S141} 

\refitem {van Amerongen, S., Damen, E., Groot, M., Kraakman, H., van Paradijs, J.} 
         {1987} {\MNRAS} {225} {93}

\refitem {Verbunt, F., Hassall, B.J.M., Pringle, J.E., Warner, B., Marang, F.} 
         {1987} {\MNRAS} {225} {113}

\refitem {Vogt, N.} {1974} {\AA} {36} {369}

\refitem {Warner, B. and O'Donoghue, D.} {1988} {\MNRAS} {233} {705}

\end {references}

\end{document}